\documentclass[iop]{emulateapj}
\usepackage{apjfonts}
\usepackage{graphicx}
\usepackage{amsmath}

\graphicspath{{./}}

\newcommand{\Msun}{\rm M_{\odot}}

\def\Omm{{\Omega_m}}

\def\Omb{{\Omega_b}}
\def\Oml{{\Omega_{\Lambda}}}

\def\beq{\begin{equation}}
\def\eeq{\end{equation}}

\newcommand{\MBH}{\rm M_{\rm{BH}}}
\newcommand{\Mstar}{\rm M_{*}}

\shorttitle{}
\shortauthors{}

\begin{document}

\title{The Co-evolution of Cosmic Entropy and Structures in the Universe}

\author
{
Xinghai Zhao \altaffilmark{1,2},
Yuexing Li \altaffilmark{1,2},
Qirong Zhu \altaffilmark{1,2} and
Derek Fox \altaffilmark{1,2}
}
\affil{$^{1}$Department of Astronomy and Astrophysics, The Pennsylvania State University,
525 Davey Lab, University Park, PA 16802, USA}

\affil{$^{2}$Institute for Gravitation and the Cosmos, The Pennsylvania State University, 104 Davey Lab, University Park, PA 16802, USA}

\email{xuz21@psu.edu}

\begin{abstract}

According to the second law of thermodynamics, the arrow of time points to an ever increasing entropy of the Universe. However, exactly how the entropy  evolves with time and what drives the growth remain largely unknown. Here, for the first time, we quantify the evolving entropy of cosmic structures using a large-scale  cosmological hydrodynamical simulation. Our simulation starts from initial conditions predicted by the leading $\Lambda$CDM cosmology, self-consistently evolves the dynamics of both dark and baryonic matter, star formation, black hole growth and feedback processes, from the cosmic dawn to the present day. Tracing the entropy contributions of these distinct components in the simulation, we find a strong link between entropy growth and structure formation. The entropy is dominated by that of the black holes in all epochs, and its evolution follows the same path as that of galaxies: it increases rapidly from a low-entropy state at high redshift until $z \sim 2$, then transits to a slower growth. Our results suggest that cosmic entropy may co-evolve with cosmic structure, and that its growth may be driven mainly by the formation of black holes in galaxies.  We predict that the entropy will continue to increase in the near future, but likely at a constant rate.  

\end{abstract}



\section{Introduction}
\label{s1}

The ``arrow of time'' in our Universe, which has enabled its evolution
to the present highly complex state, with large scale structure,
galaxies, stars, and planetary systems -- not to mention life itself
-- has its origins in the extremely low-entropy initial state of our
cosmos \citep{Penrose89,Carroll10}, and barring unforeseen calamities,
will persist for some $10^{100}$ years until the Universe achieves its
equilibrium state, aka its ``heat death''
\citep{Thomson:1852}. Throughout this extremely long yet transient
epoch, the probabilistic laws of Boltzmann's statistical mechanics
assure us, the entropy of the Universe will be always and ever
increasing, as per the Second Law of Thermodynamics. 

From this perspective, the formation and evolution of cosmic
structure, as explored via large-scale cosmological surveys and
numerical simulation, must also be always consistent with the Second
Law. As such, both the initial formation and the subsequent evolution
and (ultimately) dissolution of all the structures we observe must
contribute to the ever-increasing entropy of the Universe. 

Several studies \citep{Basu:1990,Frampton:2009,Egan:2010} have
explored the entropy budget of the present-day observable Universe. These studies suggested that today's total entropy is dominated by the entropy of the supermassive black holes (SMBHs) at the centers of galaxies. However, the detailed evolution of the entropy and its origin remain largely unknown.

Over the last several decades, numerical simulation have been very
successful on the study of the evolution of the structures in the
Universe from the time when the Universe was less than 1 million years
old to the current time at 13.7 billion years later
(e.g. \cite{Springel:2005a}). They can also trace different energy
forms in cosmic structures through vast scales, from dark matter (DM) in
the large-scale structure to gas and stars in the galaxies to ultra
compact BHs while producing results that agree well with the
astronomical observations. Thus numerical simulation is an ideal tool
to study the evolution of the entropy along the cosmic structure
formation.

In this paper, we present a quantitative study on the evolution of entropy using high-resolution cosmological
hydrodynamical simulation. In \S~2 we describe the numerical simulation, and the 
appropriate theoretical formulations for the entropy of different energy forms, using physical quantities that can be extracted directly
from the simulation. In \S~3, we present results of the formation and evolution of structures from the simulation, the evolution of cosmic entropy, and the link between entropy growth and structure formation. We summarize in \S~4 the interpretations of the results and their
implications.


\section{Methodology}

\subsection{Numerical Simulation}
\label{sim}

Our cosmological simulation includes both dark and baryonic matter and related physical processes including star formation, BH growth, and feedback. It starts from initial conditions predicted by the leading cold dark matter cosmology, $\Lambda$CDM, and follows the formation and evolution of structures from redshift $z=99$  to the current time at $z=0$.

We performed the simulation using the parallel, N-body/SPH code GADGET-3, which is an improved version of the widely used simulation code GADGET  \citep{Springel2001, Springel:2005}. GADGET uses the ``TreePM'' method to compute gravitational forces, which combines a ``tree'' algorithm for short-range forces and a Fourier transform particle-mesh method for long-range forces. It incorporates an entropy-conserving formulation of SPH with adaptive particle smoothing. Radiative cooling and heating processes are calculated assuming collisional ionization equilibrium, and the UV background model of \cite{Faucher2009} is used, which assumes that reionization was completed roughly by redshift $z \sim 6$. 

Star formation is modeled in a multi-phase ISM, with a rate that follows the Schmidt-Kennicutt Law (\citealt{Schmidt1959, Kennicutt1998}). The model of BH growth and feedback follows that of  \cite{Springel2005B} and \cite{DiMatteo2005}, where the BH accretion is calculated using a spherical Bondi rate \citep{Bondi1952} under the Eddington limit, and its feedback is in form of thermal energy, $\sim 5\%$ of the radiation, injected into surrounding gas isotropically. We follow the seeding scheme of \cite{Zhu2012} and plant a seed of mass $\rm M_{BH} = 10^{5}~ h^{-1}\,  \Msun$ in each halo once its total mass exceeds $\rm 10^{10}~ h^{-1}\, \Msun$, similar to previous cosmological simulations with BHs \citep{DiMatteo2008, DiMatteo2012}. Such a self-regulated BH model has been demonstrated to successfully reproduce many observed properties of local galaxies (e.g., \citealt{DiMatteo2005, Hopkins2006, Zhu2012}) and the most distant quasars at $z \sim 6$ \citep{Li2007}. 

The whole simulation is set up with a periodic boundary condition and the box size is 100 Mpc/h in comoving coordinates. The initial condition contains gas and DM components, each of them is represented by $512^3$ particles with a gravitational softening length $\epsilon=$ 5 kpc/h. The cosmological parameters used are $\Omm=0.27$, $\Omb=0.045$, $\Oml=0.73$ and $H_0=100$h km/s/Mpc with $\rm h=0.7$, consistent with the seven-year results of the WMAP \citep{Komatsu2011}. 

In order to identify structures formed on different scales, we use two on-the-fly group finding algorithms  in the simulation. One is the friends-of-firends (FOF) algorithm  used to link the DM particles with particle separations less than 0.2 times of the mean particle spacing. The gas, star and BH particles are then linked to the DM particle groups using the same algorithm. The other one is  the SUBFIND algorithm \citep{Springel:2001, Dolag2009} used to find gravitationally bound physical substructures in the groups. It first identifies the local overdensities using the SPH kernel interpolation, then the gravitationally unbound particles are iteratively removed. A substructure is considered physically bound if the final clumps have more than 20 particles. Throughout this work, a galaxy is defined as the group returned by the SUBFIND, which includes a DM halo, gas, stars, and BHs.


\subsection{Entropy Formulation}
\label{entropy}

We calculate the entropy of different energy forms of the Universe as follows: for gas and stars, we use the Sakur-Tetrode law \citep{Basu:1990,Egan:2010},
\begin{equation}
S_i = k N_i~ {\rm{ln}}[Z_i(T)(2\pi m_ikT)^{\frac{3}{2}}e^{\frac{5}{2}}n_i^{-1}h^{-3}],
\label{eq_s_t}
\end{equation}
where $N_i$ and $Z_i$ represent the number of particles and internal partition function of either gas or star particles, $n_i$ is the corresponding particle's number density and other symbols in Eq. (\ref{eq_s_t}) represent their normal physics quantities. For the gas particles, we take $Z$ to be 1 for general baryonic matter. This is close to the value used in \cite{Basu:1990}, but we treat different types of the baryonic particles equally. All other quantities can be extracted from the numerical simulation. For the star particles, in principle, one should use different partition functions for different stellar populations. Because of the limitation of the resolution of the current cosmological simulation, it is not yet practical to trace the evolution of each stellar population directly. So we adopt the typical value of a main sequence star and the corresponding entropy per baryon for all star particles as calculated in \cite{Basu:1990}. Since there is a simple relation between the entropy per baryon for gas and stars as used in \cite{Egan:2010}, here we use this relation and mass of the stellar component to estimate its entropy from the calculated gas entropy in the simulation.

For the entropy of DM in the Universe, it is worth noting that the nature of the DM particles is still unclear. Therefore, it is difficult to calculate the DM entropy in an exact way. In this study, we adopt the cold DM model and use a well-motivated definition from the studies of the X-ray clusters in astrophysics \citep{Navarro:1995,Eke:1998,Faltenbacher:2007}. That is, we define the ``temperature'' of a DM particle via relation,
\begin{equation}
3kT=\mu m_p \sigma^2,
\end{equation}
where $\mu$ is the mean molecular weight of the DM particle, $m_p$ is the proton mass and $\sigma$ is the three-dimensional velocity dispersion of the DM particle. $\sigma$ can be easily extracted from numerical simulation while $\mu$ is a parameter that must be constrained from current investigations of the nature of the DM particles. Here, we adopt a typical value of 40 GeV for the mass of the DM particles from recent studies of weakly interacting massive particles (WIMP) \citep{Geringer-Sameth:2011}. Now that we have a definition of the ``temperature'' of a DM particle, we can use Eq. (\ref{eq_s_t}) to calculate the entropy of the DM component in a way similar to that of the gas component in the simulation. We want to emphasize that this method is only to estimate the DM entropy from an astrophysics perspective.

To calculate the entropy of BHs, we use the Bekenstein-Hawking formula \citep{Bekenstein:1973, Hawking:1976},
\begin{equation}
S_{BH}=\frac{kc^3A}{4G\hbar},
\end{equation}
where $A$ is the surface area of a BH. For simplicity, we treat the BHs in the simulation as non-rotating charge-free Schwarzschild BHs. So the surface area of the BHs can be calculated as $A=16\pi G^2M^2/c^4$, where $M$ is the BH mass. Then the entropy of a BH particle in the simulation can be calculated as,
\begin{equation}
S_{BH}=k\frac{4\pi G}{c\hbar}M^2.
\label{eq_b_h}
\end{equation}

As we can see from Eq. (\ref{eq_b_h}), the BH entropy is proportional to the square of its mass. This leads to a BH entropy dominant era in the cosmic evolution as we will illustrate in our results. But we need to keep in mind that the derivation of the BH entropy is not from the same origin as that of gas, star and DM. Its exact physical interpretation is still a matter of debate.


\section{Results}
\label{result}

\subsection{Formation and Evolution of Cosmic Structures}

\begin{figure*}
\begin{center}
\includegraphics[trim=0cm 0cm 0cm 0cm, clip=true, angle=0, width=2.1in]{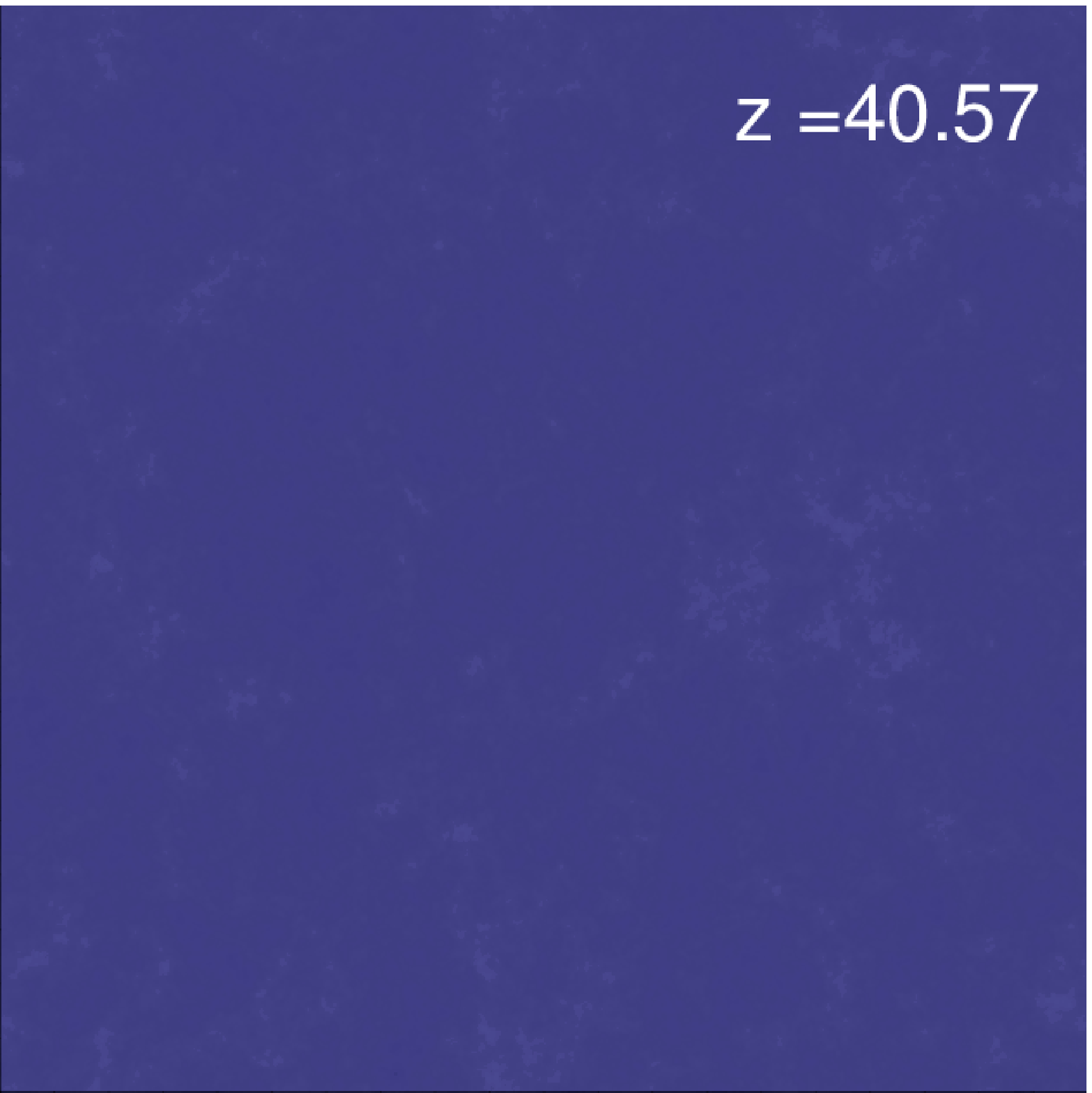}
\includegraphics[trim=0cm 0cm 0cm 0cm, clip=true, angle=0, width=2.1in]{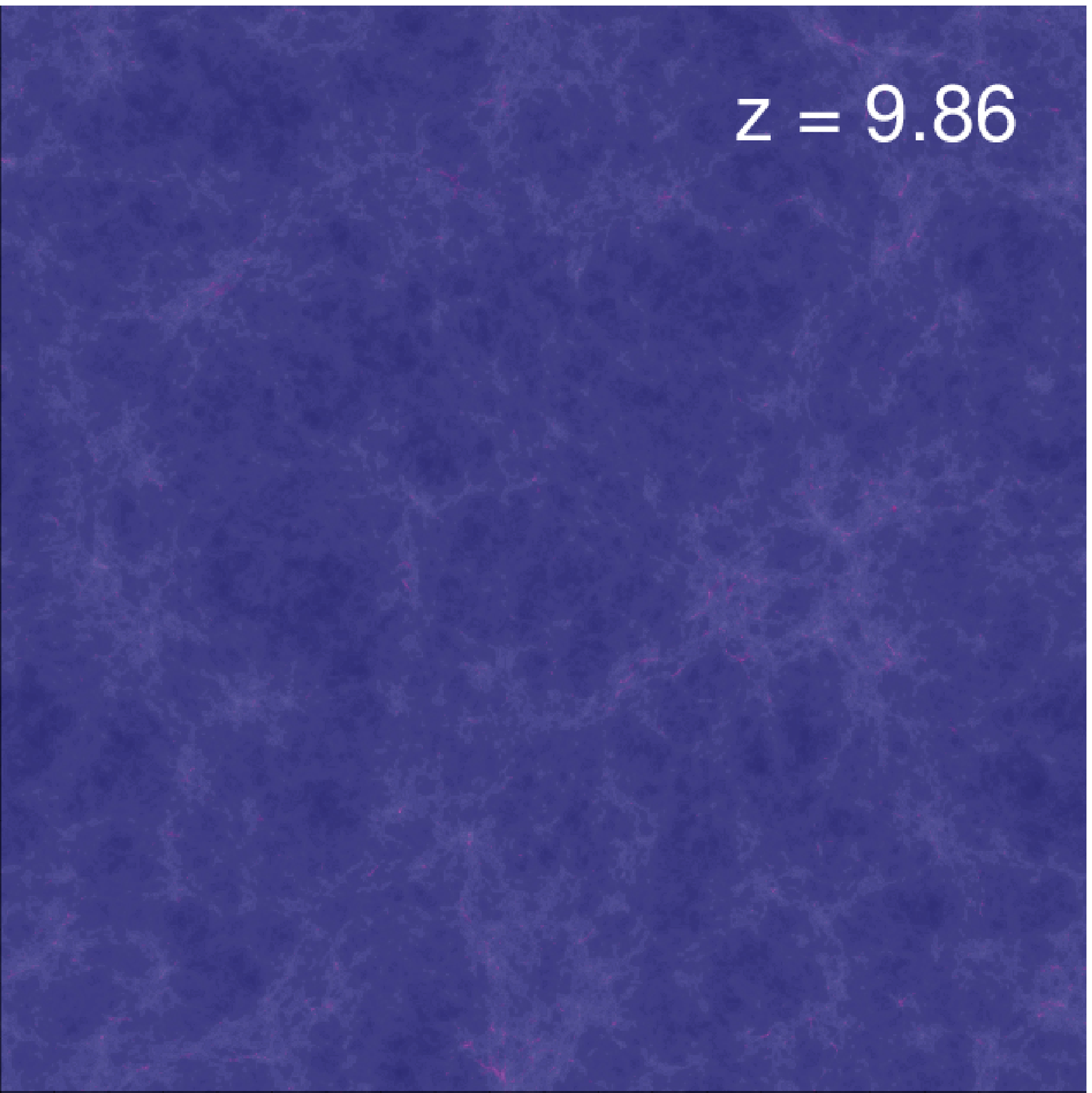} 
\includegraphics[trim=0cm 0cm 0cm 0cm, clip=true, angle=0, width=2.1in]{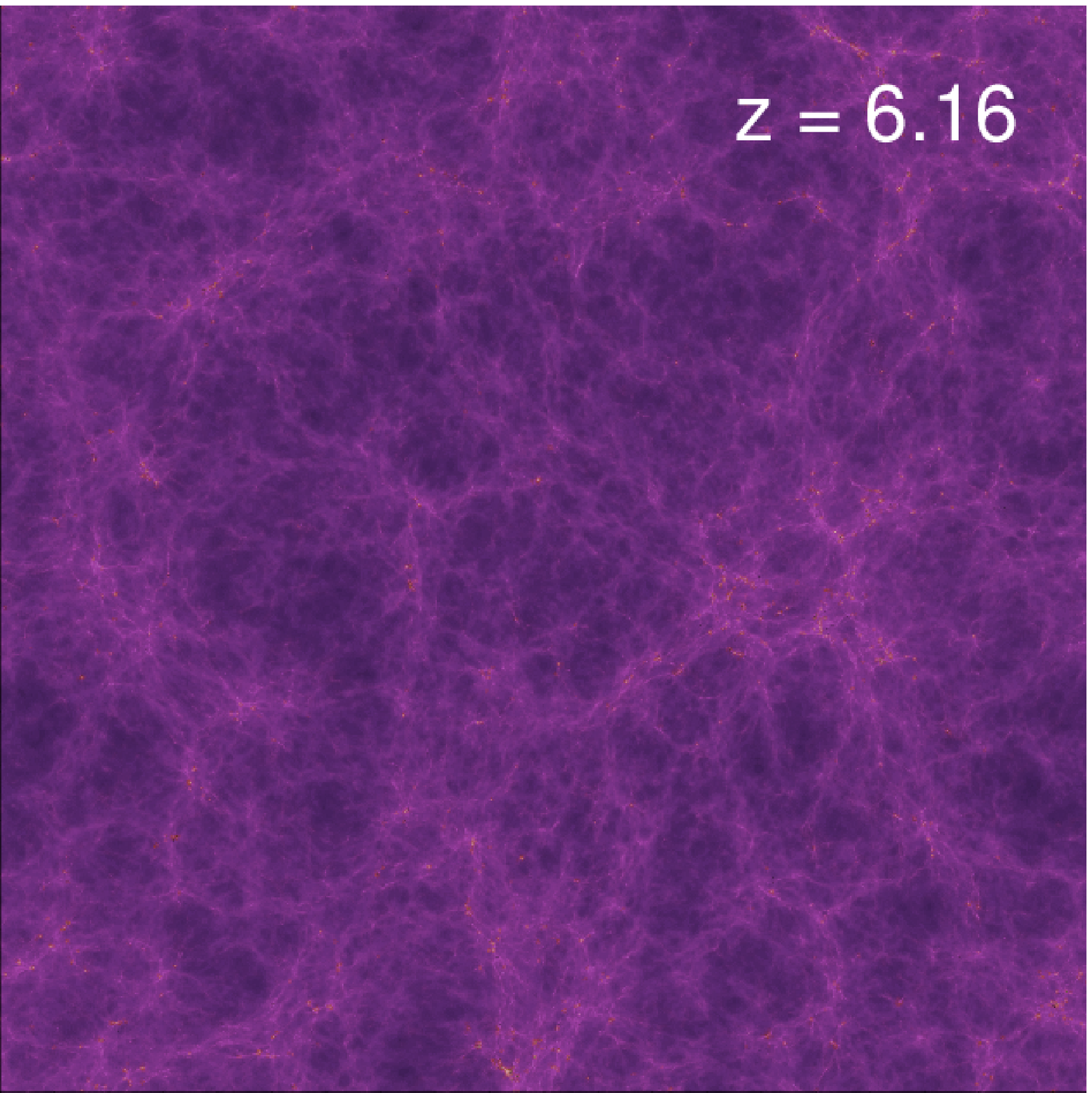} \\
\includegraphics[trim=0cm 0cm 0cm 0cm, clip=true, angle=0, width=2.1in]{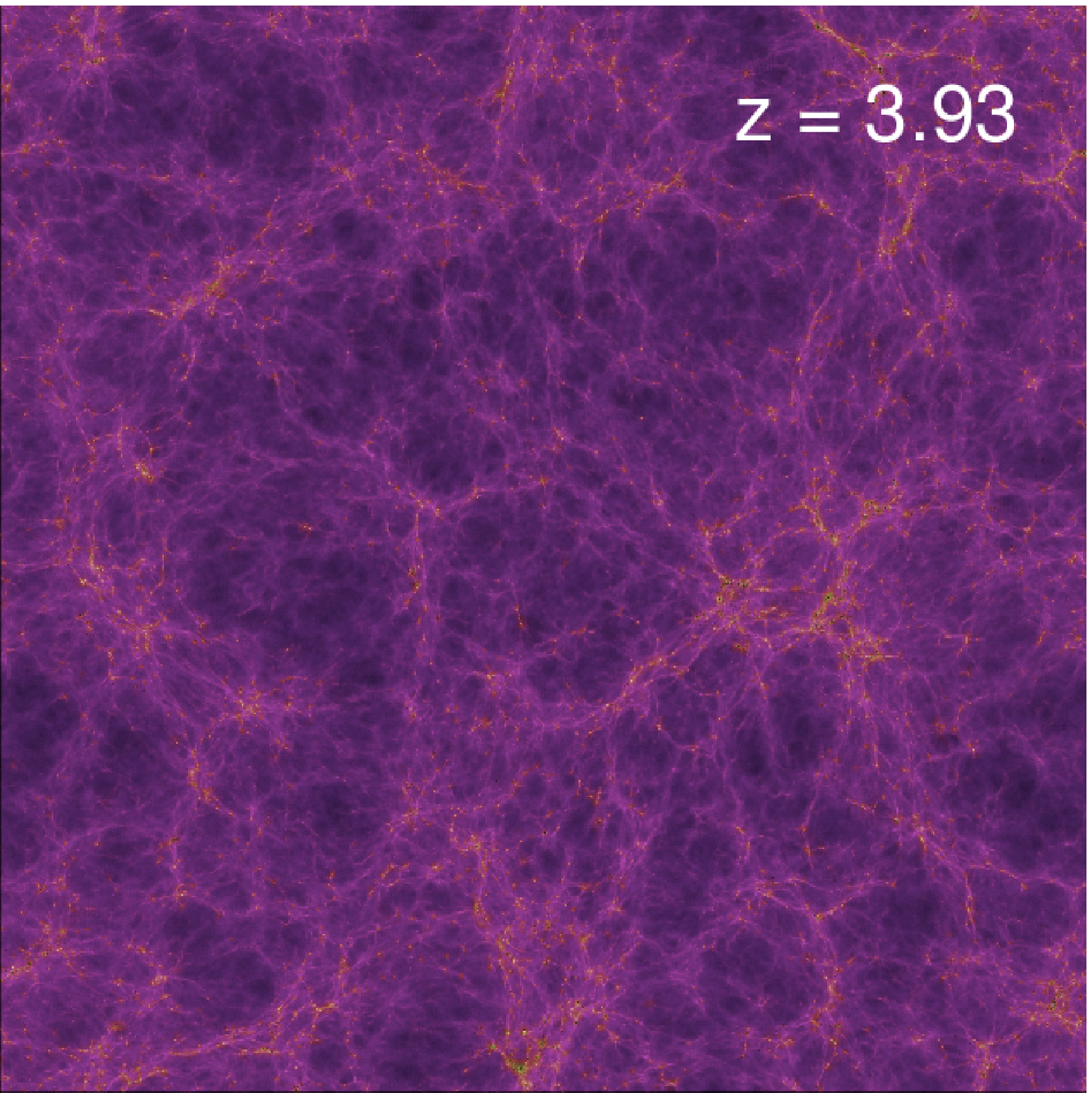} 
\includegraphics[trim=0cm 0cm 0cm 0cm, clip=true, angle=0, width=2.1in]{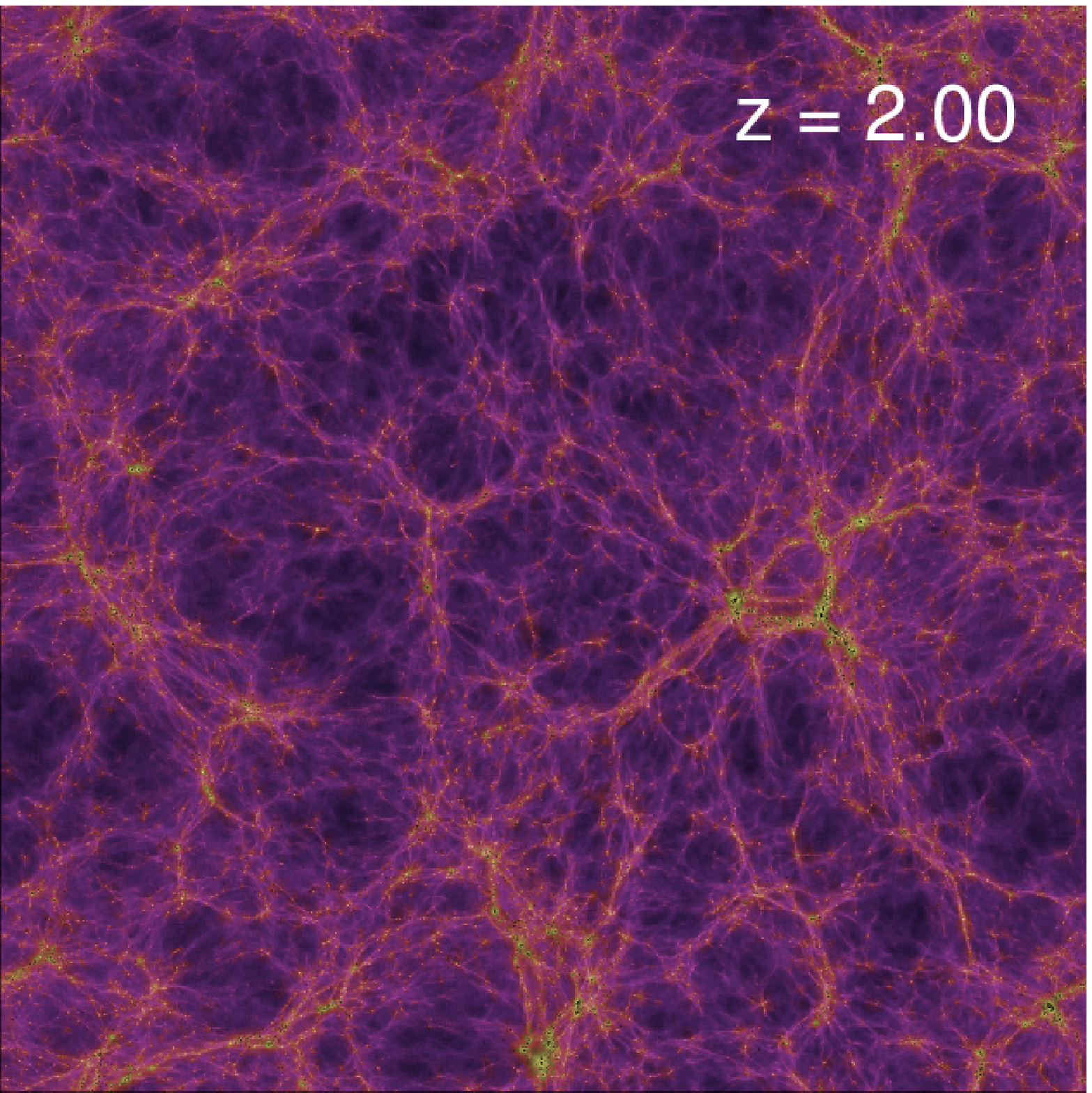}
\includegraphics[trim=0cm 0cm 0cm 0cm, clip=true, angle=0, width=2.1in]{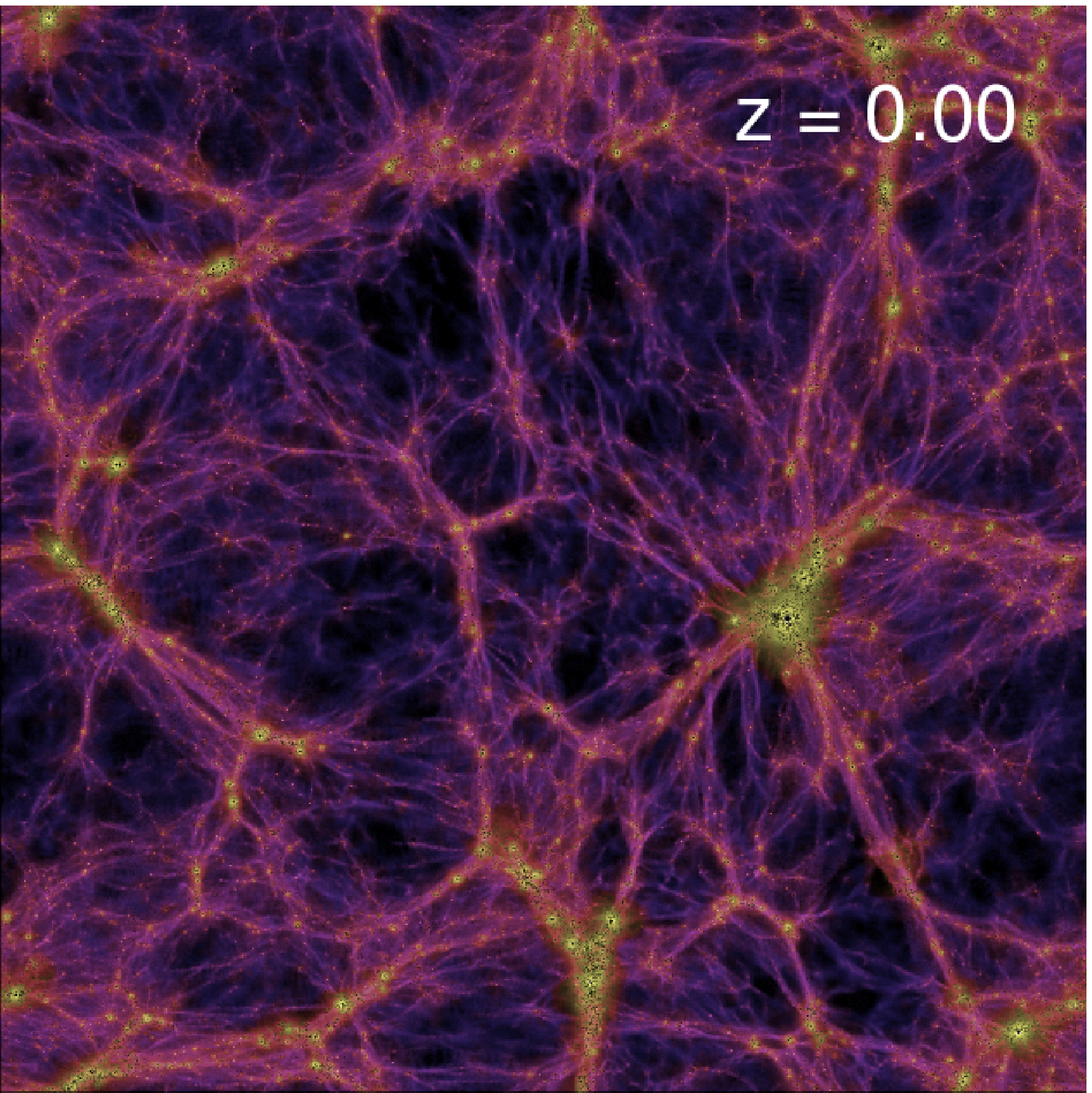}
\caption{Formation and evolution of structures from the cosmological simulation. The images show the 2-D projected density of both gas and stellar components in a spatial slice in Z direction with a thickness of 10 Mpc/h (comoving). For the gas and stars, the brightness corresponds to the density while the color corresponds to the temperature of the gas (blue indicates cold gas, brown indicates hot, tenuous  gas) and the metallicity of the stars (in yellow color). The BHs are represented by the black  dots, the size of which is proportional to the BH mass. The box size is 100 $h^{-1} $Mpc in comoving coordinates.}
\label{fg_snap}
\end{center}
\end{figure*}

\begin{figure}
\begin{center}
\includegraphics[angle=0, width=3.6in]{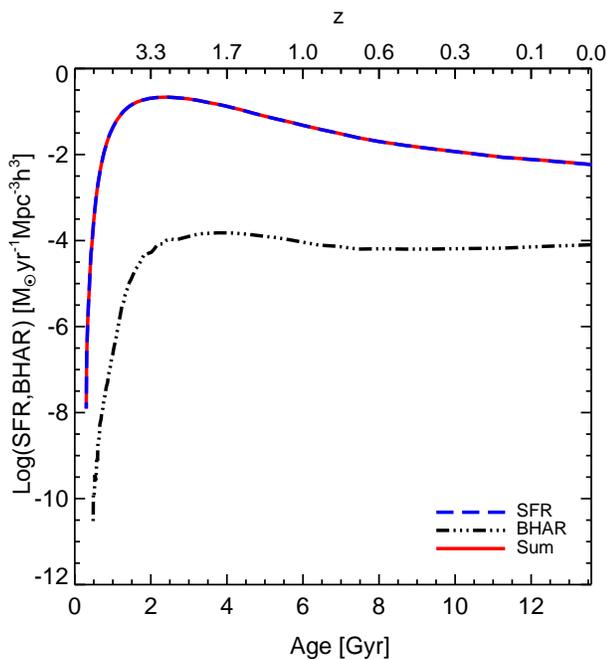}
\caption{The growth history of structures from the simulation, as indicated by the density of the star formation rate (blue  line), the BH accretion rate (black line), and their sum (red line) as a function of time.}
\label{fg_growth}
\end{center}
\end{figure}

Owing to the formation of structures, the Universe evolves from a nearly homogeneous distribution of matter to a filamentary complex, the cosmic webs, with time, as demonstrated in Figure~\ref{fg_snap}. Dense regions build up gradually along the filaments due to density fluctuation and gravitational instability, DM and gas in these regions collapse to form galaxies. These galaxies interact with each other and merge hierarchically to form ever larger ones. The BHs form in these massive halos, and they grow through gas accretion and mergers following the hierarchical buildup of their host galaxies. At high redshifts ($z \gtrsim 10$), the gas and DM are the dominant components, but stars and BHs form rapidly at later time ($z < 10$).  

The growth history of galaxies and BHs are shown in Figure~\ref{fg_growth}. The star formation rate (SFR) rises sharply about 0.5 Gyrs ($z \sim 10$) after the Big Bang, it reaches its peak at around 2 Gyr  ($z \sim 3$), then declines gradually with time. The BH accretion lags behind the star formation slightly, but it shares similar trend. It increases rapidly to its peak at $z \sim 2$, after that it decreases slowly and maintains a constant rate at late time. As a result of such growth, the accumulated mass of the galaxies,  stars, and BHs increase monotonically with time. The BH accretion is about two to three orders of magnitude lower than the SFR, which may explain the tight correlation between the masses of the SMBHs and the stellar masses of their host galaxies, $\MBH \sim 10{-2} - 10^{-3}\, \Mstar$ observed in nearby galaxies \citep{Magorrian1998, Haring2004}.

Figure~\ref{fg_growth} clearly shows that the growth of the cosmic structure is driven by star formation at early time and BH formation at late time.The similarity between the star formation and BH growth indicates a co-evolution between galaxies and BHs. 

\subsection{Evolution of Cosmic Entropy}

\begin{figure}
\begin{center}
\includegraphics[angle=0, width=3.4in]{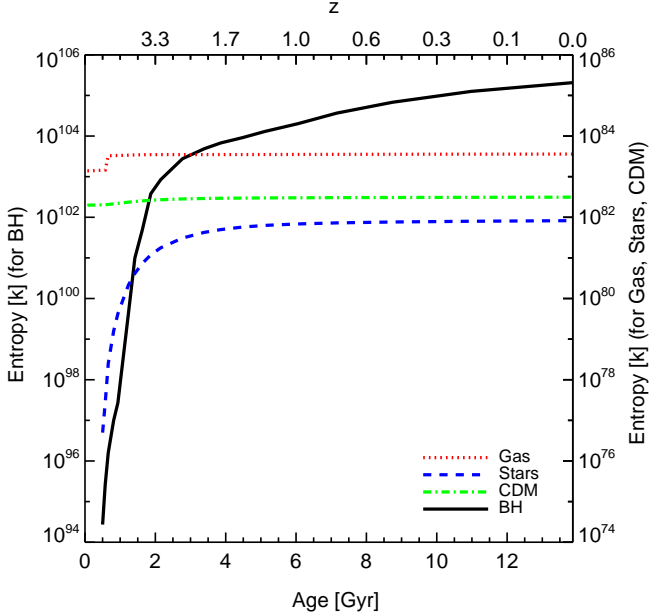}
\caption{The evolution history of the entropy of the gas, stars, DM and BHs in the Universe. Note the different scales on the left and right axises. The left y-axis corresponds to the BH entropy, while the right y-axis corresponds to the gas, star and DM entropy.}
\label{fg_entropy}
\end{center}
\end{figure}

Using the formulation described in Sec. \ref{entropy}, we calculate the evolution history of the entropy of different energy forms in the observable Universe from redshift $z=41$ to $z=0$. In order to calculate the entropy of the whole observable universe, we first calculate it using our simulation data with a simulation box size of 100 Mpc/h and then scale this result up to the size of the observable universe with a typical value of 14.3 Gpc in radius \citep{Cornish:2004,Key:2007,Egan:2010,Bielewicz:2011} using the fact that the Universe is very close to uniform on the scales larger than 100 Mpc/h. The resulting entropy of different structure components are shown in Figure~\ref{fg_entropy}. 

For gas (baryonic matter), the entropy only has a very slight increase in the redshift range studied except  a jump at redshift $8 < z < 9$ because of the cosmic reionization that is implemented in the simulation. The cosmic entropy at present day in this study is, however, about one magnitude higher than the estimation in \cite{Egan:2010} as shown in Table~\ref{tab_entropy_comp}. We believe this is caused by the fact that we can calculate the temperature of individual gas particles in the simulation while \cite{Egan:2010} only uses the estimation of the whole gas component. Thus, our calculation is likely to be more accurate than the previous studies.

For the entropy of the stellar component, since it is calculated by scaling the entropy of the gas with their relative mass ratio, its present day value is also higher than the one in the previous studies \citep{Egan:2010}. But, unlike the gas component, the stellar component's entropy increases dramatically once stars form in the simulation until redshift $z \sim 2$. Then it gradually increases to the present day's value. This is due to the rapid cosmic star formation before $z \sim 2$.

For the entropy of the DM, the way we calculate the entropy is different from the one in \cite{Egan:2010} in which the relativistic degrees of freedom is used. So it is not straightforward to make a comparison in this case. Similar to the gas entropy, the DM entropy also increases slightly during the structure formation process.

For the entropy of the BHs, due to the limitation of the resolution of the simulation, we can only include the BHs with masses larger than $\rm 10^{5}~ h^{-1} \, \Msun$. So the BH entropy value we calculate in this study corresponds to the entropy of the SMBHs in \cite{Egan:2010}. Taking account that there are likely more BHs with masses close to $10^5\, \Msun$ in nature, the actual value of the BH entropy may be even higher.
 
The resulting entropy of BHs is at least more than 20 orders of magnitude higher than the other energy forms throughout the cosmic time in Figure~\ref{fg_entropy}.  Therefore, the total entropy of the Universe is dominated by that of the BHs. This is consistent with conclusions from previous studies for the present-day Universe \citep{Basu:1990,Frampton:2009,Egan:2010}.

Table~\ref{tab_entropy_comp} gives a comparison of the different estimates of the entropy of different energy forms in the local observable universe. It shows that our simulation results are close to those from previous studies.  

From Figure~\ref{fg_entropy}, we clearly see that the growth of the cosmic entropy is not a simple linear curve. It increases rapidly from a low-entropy state after the Big Bang until $z \sim 2$, then transits to a slower growth. Unlike the entropies of the stars and gas which only shows significant increase at early time but flattens at later times, the entropy of the BHs keeps increasing with time. This trend is likely to continue in the near future. 

\begin{deluxetable}{ccc}
\tablecaption{A comparison of different estimates of entropy of different energy forms in the present-day observable universe. 
\label{tab_entropy_comp}}
\tablehead{
\colhead{Energy form} & 
\colhead{Entropy (this study)} &
\colhead{Entropy (previous studies)\tablenotemark{a}}
\vspace{0.05cm} \\
&
\colhead{[$k$]} &
\colhead{[$k$]}
}  
\startdata
BHs & $2.06 \times 10^{105}$ & $10^{106}[1], 10^{102}[2]$\\
Gas & $3.60 \times 10^{83}$  & $10^{82}$[1]\\
Stars & $8.25 \times 10^{81}$ & $10^{81}$[1], $10^{79}$[2]\\
CDM & $3.13 \times 10^{82}$  & -\tablenotemark{b}
\enddata 
\tablenotetext{a}{[1] \cite{Egan:2010}, [2] \cite{Frampton:2009}}
\tablenotetext{b}{The way we calculate the entropy is different from the one in \cite{Egan:2010} in which the relativistic degrees of freedom is used. So it is not straightforward to make a comparison in this case.}
\end{deluxetable}

\subsection{Structure Formation as Driver of Entropy Growth}

\begin{figure}
\begin{center}
\includegraphics[angle=0, width=3.4in]{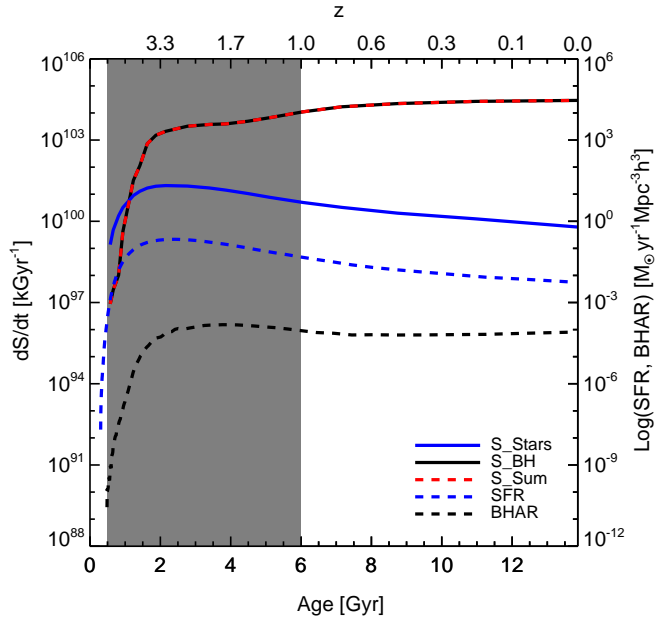}
\caption{The evolution history of the entropy growth rate of BHs (black solid line), stars (blue solid line), and their sum (red dashed line), contrasted by the star formation rate (blue dashed line) and BH accretion rate (black dashed line). Note the difference in the y-axises. The left y-axis corresponds to the entropy growth rate, while the right y-axis corresponds to the structure growth rates. The shaded region indicates the period during which the entropy and structure grow rapidly and simultaneously.}
\label{fg_dev_growth}
\end{center}
\end{figure}

The evolution of entropy in Figure~\ref{fg_entropy} and the structure formation history in Figure~\ref{fg_growth} show a strikingly similar trend, as both increase rapidly at about the same time. This hints a possible link between the two.

To investigate the link between entropy and structure formation, we plot the growth rate of entropy as a function of time, in comparison with the SFR and BH accretion rate, as shown in Figure~\ref{fg_dev_growth}. It clearly shows that the the generation of entropy is closely related to the SFR and the BH accretion rate over the redshift range studied. This coincidence suggests that cosmic entropy growth is driven by structure formation.



\section{CONCLUSION}
\label{summary}

We have used a large-scale cosmological simulation to study the evolution history of the cosmic entropy along the structure formation in the Universe. Our calculations are based on well-motivated physical formulations using parameters directly from the simulation.  Our estimations of the entropy of different energy forms in the local observable Universe are in broad agreement with previous studies. 

Moreover, by tracking the evolution of entropy of different energy forms, we find that the entropy of BHs is at least 20 orders of magnitude higher than the other components including gas, stars and DM. Therefore, the cosmic entropy is dominated by that of the BHs. Rather than follow a simple linear curve, it has roughly two distinctive growth phases: a rapid increases phase from a low-entropy state at high redshift until $z \sim 2$, then transits to a slower growth phase. 

Furthermore, we find a strikingly similar evolution between the growth rate of entropy and that of galaxies. This suggests that the cosmic entropy co-evolves with cosmic structures, and that generation of cosmic entropy is driven by structure formation in the Universe. We predict that the BH entropy, and hence the cosmic entropy, will continue to increase in the near future at a constant rate. 

We note that due to the limitation of the resolution of the simulation, we can only include the BHs with masses larger than $\rm 10^{5}~ h^{-1} \, \Msun$. Taking into account smaller BHs may increase the actual value of the BH entropy. In addition, due to the unknown nature of DM, our calculation of its entropy may be subject to a large uncertainty. Nevertheless, these uncertainties would not affect our conclusion that the structure formation is the main driving force in the generation and evolution of the cosmic entropy.


\section*{Acknowledgments}

Support from NSF grants AST-0965694 and AST-1009867 is gratefully acknowledged. We acknowledge the Research Computing and Cyberinfrastructure unit of Information Technology Services at the Pennsylvania State University for providing computational resources and services that have contributed to the research results reported in this paper (URL: http://rcc.its.psu.edu). The Institute for Gravitation and the Cosmos is supported by the Eberly College of Science and the Office of the Senior Vice President for Research at the Pennsylvania State University.



\end{document}